\documentclass[11pt]{article}
\input{epsf}
\usepackage{epsfig,amsmath,amssymb}

\begin{document}

 \title{
T-duality in Ramond-Ramond backgrounds}

 \author{Raphael Benichou, Giuseppe Policastro, Jan Troost   }
 
\maketitle

\begin{center}
 \emph{Laboratoire de Physique Th\'eorique} \footnote{Unit\'e Mixte du CRNS et
    de l'Ecole Normale Sup\'erieure associ\'ee \`a l'universit\'e Pierre et
    Marie Curie 6, UMR
    8549. Preprint LPTENS-08/03.},
\emph{ Ecole Normale Sup\'erieure,  \\
24 rue Lhomond, F--75231 Paris Cedex 05, France}
\end{center}

 \abstract{Using the pure spinor formalism on the world-sheet, we derive the
   T-duality rules for all target space couplings in an efficient
   manner. The world-sheet path integral derivation is a proof of the equivalence of the
   T-dual Ramond-Ramond backgrounds which is valid 
non-perturbatively in the string length over the curvature
   radius and to all orders in perturbation theory in the string coupling.}

\section{Introduction}
Target space duality is a symmetry of string theory that maps a string theory
in a background to a dual string theory in a dual background. For reviews see
\cite{Alvarez:1994dn}\cite{Giveon:1994fu}.  The map between T-dual backgrounds
was derived in a world-sheet path integral formalism for Neveu-Schwarz
Neveu-Schwarz fields in \cite{Buscher:1987qj}.  For Ramond-Ramond backgrounds
in type II, various derivations have appeared in the literature: the authors
of \cite{Bergshoeff:1995as} used the equivalence of the type II supergravity
actions after reduction to nine dimensions. In \cite{Hassan:1999bv} arguments
were given for the
transformation of the spacetime supersymmetry parameters, and
then the transformations of the gravitini and of the Ramond-Ramond fields were inferred
by demanding compatibility between T-duality and supersymmetry.  A world-sheet
derivation was obtained in the Green-Schwarz formalism in \cite{Cvetic:1999zs}
up to quadratic order in the superspace coordinate, and later extended to all
orders in \cite{Kulik:2000nr}\cite{Bandos:2003bz}.

In the present letter, we give a novel world-sheet derivation of T-duality
based on the pure spinor formalism \cite{Berkovits:2000fe}. Our motivation for
revisiting this problem is twofold: first, the duality rules are derived in a
simpler and more streamlined way than with other methods.  Secondly, since the
pure spinor formalism gives a satisfactory conformal field theory
 description of the string in generic
backgrounds, we are able to promote the duality to the path integral level,
thus providing for a derivation of T-duality which is valid non-perturbatively
in the string length over the curvature radius and to all orders in the string
coupling. The duality is valid in the presence of Ramond-Ramond and fermionic
backgrounds.

\section{Derivation of the classical T-duality rules}

We will derive the T-duality rules from the world-sheet pure spinor
formalism \cite{Berkovits:2000fe}. (See \cite{Berkovits:2002zk} for
a review.) The derivation will have the advantage of simplicity, and the
formalism is
suitable for the full quantum theory since in the pure spinor formalism the 
theory can be quantized without obstruction. 

The pure spinor world-sheet action is :
\begin{eqnarray} \label{action}
S &=& \frac{1}{2 \pi \alpha '} \int d^2 z \left[
\frac{1}{2}(G_{MN}(Z)+B_{MN}(Z))\partial Z^M \bar{\partial} Z^N +  P^{\alpha \hat{\beta}}(Z) d_{\alpha} \hat{d}_{\hat{\beta}}\right. \nonumber \\
  & & +\ E^{\alpha}_M (Z) d_{\alpha} \bar{\partial}Z^M +
E^{\hat{\alpha}}_M (Z) \hat{d}_{\hat{\alpha}} \partial Z^M + \Omega_{M \alpha}\,^{\beta}(Z) \lambda^{\alpha} w_{\beta} \bar{\partial}Z^M + 
\hat{\Omega}_{M \hat{\alpha}}\,^{\hat{\beta}}(Z) \hat{\lambda}^{\hat{\alpha}}
\hat{w}_{\hat{\beta}} \partial Z^M \nonumber \\
  & & \left. +\ C_{\alpha}^{\beta \hat{\gamma}}(Z) \lambda^{\alpha} w_{\beta}
  \hat{d}_{\hat{\gamma}} + \hat{C}_{\hat{\alpha}}^{\hat{\beta} \gamma}(Z)
  \hat{\lambda}^{\hat{\alpha}} \hat{w}_{\hat{\beta}} d_{\gamma} + S_{\alpha \hat{\gamma}}^{\beta \hat{\delta}}(Z) \lambda^{\alpha} w_{\beta}
  \hat{\lambda}^{\hat{\gamma}} \hat{w}_{\hat{\delta}} \right]  \nonumber \\
  &+ &\frac{1}{4 \pi} \int d^2 z (\Phi(Z) {\cal R}^{(2)}) + S_{\lambda} +
    \hat{S}_{\hat{\lambda}} 
\end{eqnarray}
where the coordinates 
$Z^M=(x^\mu,\theta^\alpha,\hat{\theta}^{\hat{\alpha}})$ are coordinates
on an $\mathbb{R}^{10|32}$ superspace. The variables $d$ and $\hat{d}$ as well as
$w$ and $\hat{w}$  are independent
spinorial variables while the variables $\lambda$ and $\hat{\lambda}$ are pure
spinors, which means that they are Weyl spinors satisfying the constraint 
\begin{equation}\label{constr}
\lambda^{\alpha} \gamma^{a}_{\alpha \beta} \lambda^{\beta} = 0 \,. 
\end{equation}
The action for the pure spinors, $S_{\lambda}+\hat S_{\hat\lambda}$, is formally
a free field action. Because of the constraints a proper treatment requires
care. That will not be important for our purposes.

The action (\ref{action}) describes both the type IIB and IIA string. The only
difference is whether the hatted and un-hatted spinor indices have the same or
the opposite chirality.  All the couplings are superfields, which means they
are generic functions of all the superspace coordinates.  For the reader's
convenience, we recall the meaning of the various superfields : $G_{MN}$ is
the metric, $B_{MN}$ is the B-field, $P^{\alpha \hat \beta}$ are the RR field
strengths, $E_{M}^{\alpha}$ is the spinorial part of the vielbein,
$\Omega_{M\alpha}^{~~~\beta} $ is the spin connection, $C_{\alpha}^{\beta \hat
  \gamma}$ contains the field strength of the dilatino, $S_{\alpha \hat
  \gamma}^{\beta \hat \delta} $ contains the Riemann curvature, and $\Phi$ is
the dilaton.  The world-sheet curvature is denoted ${\cal R}^{(2)}$, and it
couples to the dilaton via the Fradkin-Tseytlin term in the last
line. Superdiffeomorphisms and local Lorentz transformations
 allow to make particular gauge choices where the physical content of the
fields is more manifest. For brevity's sake, we refer again to
\cite{Berkovits:2002zk}.

As mentioned in \cite{Berkovits:2002zk} and explained in detail in \cite{Berkovits:2001ue}, 
the action enjoys a BRST symmetry generated by the charge 
$$Q = \int (\lambda^{\alpha} d_{\alpha} + \hat \lambda^{\hat\alpha} \hat
d_{\hat \alpha} )$$
when the background fields satisfy a set of constraints,
which are known to put all the fields on-shell.
 
For our purposes it is important to notice that the action (\ref{action}) is
the most general one that respects the local symmetries. 
Since, as it turns out, the BRST operator will not be affected by
the T-duality, it follows immediately that the T-duality transformed fields
will again solve the equations of motion. This had to be imposed as a
requirement in order to find the form of the transformations in the
Green-Schwarz formalism \cite{Kulik:2000nr}, but in the pure spinor formalism
it is automatically true.

The difference with the Green-Schwarz string in this respect is 
due to the fact that the action (\ref{action}) explicitly contains all the
fields that appear in the superspace description of the gravity multiplet. The
Green-Schwarz action instead depends explicitly only on the bosonic part of
the supervielbein, so that additional input is required in order to find the
transformation rules for the other fields. In the pure spinor formalism, when
we perform the T-duality transformation on the action, we are bound to find an
action of the same form as (\ref{action}), and we can directly read off the
transformation rules for all superfields.

For now we concentrate on the first three lines of the action (\ref{action})
-- we come back to the last line later on.  We suppose that a Killing vector
exists in space-time, and we choose local coordinates such that all space-time
superfields do not depend on the coordinate $x^1$. The tangent vector
$\partial/\partial x^1$ is proportional to the Killing vector in a given
patch. The action then has a global shift symmetry (in $x^1$) that we gauge by
introducing a gauge field one-form $A$ on the world-sheet.  We moreover add a
term to the action that corresponds to a Lagrange multiplier $y$ multiplying
the field strength $F=dA$ :
 \begin{equation}\label{gaugeact}
 S_{gauged}[\partial x_{1}, A, y] = S[\partial x^{1} - A] +
\frac{1}{2 \pi \alpha'}
 \int d^{2}z \, y \, (\partial \bar A - \bar \partial A )
 \end{equation}
The model we obtain in this way is locally and classically equivalent to the
original model, since the equation of motion for the Lagrange multiplier
forces the gauge field to be pure gauge \cite{Buscher:1987qj}. The original dynamics survives unaltered.

On the other hand, we can gauge $x^1$ to zero, and we obtain an action that is
quadratic in the gauge field $A$. It is possible to integrate out the gauge
field classically, but the Gaussian integral to be performed requires a
regularization, that we discuss in the following section.  The naive
integration results in the dual action:

\begin{eqnarray}\label{dualact}
& & S = \frac{1}{2 \pi \alpha'} \int d^2 z \left\{ \begin{array}{c} \\
    \frac{}{} \\ \end{array} \right. \nonumber
\\
& & \frac{1}{2} \left[ 
   \left( \frac{4}{G_{11}} \right) (\partial y \bar{\partial} y) 
  + \left( - 2 \frac{G_{1M}+B_{1M}}{G_{11}} \right) (\partial y \bar{\partial} Z^M) 
  + \left( 2 \frac{G_{M1}+B_{M1}}{G_{11}} \right) (\bar{\partial} y \partial
   Z^M) 
\right. \nonumber \\
&& \left.  + \left( G_{MN}+B_{MN} - (-)^{MN} \frac{(G_{M1}+B_{M1})(G_{1N}+B_{1N})}{G_{11}}
   \right) (\partial Z^M \bar{\partial} Z^N)
\right] \nonumber \\
&& + \left( P^{\alpha \hat{\beta}} + \frac{2 E^{\alpha}_1 E^{\hat{\beta}}_1}{G_{11}} \right) d_{\alpha}
   \hat{d}_{\hat{\beta}} \nonumber \\
&& + \frac{2 E^{\alpha}_1}{G_{11}} d_{\alpha} \bar{\partial}y
  + \left( E^{\alpha}_M - \frac{(G_{1M}+B_{1M}) E^{\alpha}_1}{G_{11}} \right) d_{\alpha}
   \bar{\partial}Z^M \nonumber \\
&& - \frac{2 E^{\hat{\beta}}_1}{G_{11}} \partial y \hat{d}_{\hat{\beta}}
  + \left( E^{\hat{\beta}}_M - \frac{(G_{M1}+B_{M1})
   E^{\hat{\beta}}_1}{G_{11}} \right)
   \hat{d}_{\hat{\beta}} \partial Z^M  \\
&& + \frac{2 \Omega_{1 \alpha}\,^{\beta}}{G_{11}} \lambda^{\alpha} w_{\beta}
   \bar{\partial} y 
   + \left(- \frac{G_{1M}+B_{1M}}{G_{11}}\Omega_{1 \alpha}\,^{\beta} + \Omega_{M
   \alpha}\,^{\beta} \right) \lambda^{\alpha} w_{\beta}
   \bar{\partial}Z^M \nonumber \\
&& - \frac{2 \hat{\Omega}_{1 \hat{\alpha}}\,^{\hat{\beta}}}{G_{11}}
   \hat{\lambda}^{\hat{\alpha}} \hat{w}_{\hat{\beta}} \partial y 
   + \left( - \frac{G_{M1}+B_{M1}}{G_{11}}\hat{\Omega}_{1 \hat{\alpha}}\,^{\hat{\beta}} + \hat{\Omega}_{M
   \hat{\alpha}}\,^{\hat{\beta}} \right) \hat{\lambda}^{\hat{\alpha}} \hat{w}_{\hat{\beta}} \partial Z^M \nonumber
   \\
&& + \left( C_{\alpha}^{\beta \hat{\gamma}} - 
\frac{2}{G_{11}}
\Omega_{1 \alpha}\,^{\beta}
   E_1^{\hat{\gamma}} \right) \lambda^{\alpha} w_{\beta} \hat{d}_{\hat{\gamma}} 
   + \left( \hat{C}_{\hat{\alpha}}^{\hat{\beta} \gamma} - 
\frac{2}{G_{11}}
 \hat{\Omega}_{1
   \hat{\alpha}}\,^{\hat{\beta}} E_1^{\gamma} \right) \hat{\lambda}^{\hat{\alpha}}
   \hat{w}_{\hat{\beta}} d_{\gamma} \nonumber \\
& &  \left. + \left( S_{\alpha \hat{\alpha}}^{\beta \hat{\beta}}
 - 
\frac{2}{G_{11}}
\Omega_{1
   \alpha}\,^{\beta} \hat{\Omega}_{1 \hat{\alpha}}\,^{\hat{\beta}} \right) \lambda^{\alpha} w_{\beta} \hat{\lambda}^{\hat{\alpha}}
   \hat{w}_{\hat{\beta}} \right\}. \nonumber 
\end{eqnarray}
In the action (\ref{dualact}) the sum over the $M$ index is now over all variables except
$x^1$. The sign $(-)^{MN}$ is $-1$ when $M$ and $N$ are fermionic indices and
$+1$ otherwise.
It should be clear that the first two lines give the classical T-duality rules for the
NSNS sector background fields. To proceed, we take a closer look at the fourth
and fifth lines that code the transformation properties of the fermionic
vielbeins. Given the matrices $Q$ and $\hat{Q}$:
\begin{eqnarray}
{Q_M}^N &=  & \left( \begin{array}{cc}
\frac{2}{G_{11}} &  0_{1 \times 9|32}              \\
       - \frac{1}{G_{11}} (G_{1M}+B_{1M})  & 1_{9|32 \times 9|32}

\end{array} \right) 
\nonumber \\
{{\hat{Q}_{M}}}^{\, \, \, \, \, \, N} &  =  & \left( \begin{array}{cc}
-\frac{2}{G_{11}} &    0_{1 \times 9|32}          \\
        - \frac{1}{G_{11}} (G_{M1}+B_{M1})   & 1_{9|32 \times 9|32}
\end{array} \right) 
\end{eqnarray}
we see that the supervielbeins transform as $E_{M}^{'~\alpha} = {Q_{M}}^{N}\,
E_{N}^{~\alpha}$ and ${{\hat E}_{M}}^{' \hat \alpha} = \hat{Q}_{M}{}^{N} \hat{E}_{N}^{~\hat \alpha}$.  
From the action we cannot directly infer the transformation rule for the bosonic vielbein, 
since it only appears via the metric $G_{MN} = E_{M}^{a} E_{N}^{b}
\eta_{ab}$. Either $Q$ or $\hat Q$ acting on $E_{M}^{a}$ gives a rule
compatible with the transformation of the metric. So there are two candidates
for a T-dual vielbein.   
We note (similarly as in
\cite{Hassan:1999bv}) that the two possibilities are related by the transformation $Q {\hat{Q}}^{-1}$ 
which is a Lorentz
transformation of determinant -1. In fact it is a parity transformation in the direction of the T-duality. 
In order to be consistent, we have to act with a parity transformation in one spinor sector. 
We can choose it to be in the hatted sector, the other
choice being entirely equivalent.  The chiral change in parity takes us from a
type IIA/B background to a type IIB/A background.  We can then put the action
in the original form by redefining the right-moving fermions as follows:
\begin{eqnarray}
\hat \psi' = \Gamma \hat \psi 
\end{eqnarray}
where $\hat \psi = \hat \lambda, \hat \theta, \hat w, \hat d$ is any of the
spinorial fields, and $\Gamma = \gamma_{1} $.  It can be checked that after
the redefinition of the fermionic variables the action is indeed of the
original form, but the background superfields with hatted spinor indices
have to be redefined. Moreover, since $\Gamma^{2} = 1$, the
T-duality rules leave the BRST charge invariant, as anticipated.  We can now
give the transformation rules for all the superfields.  Combining the metric
and B-field in the tensor $L_{MN} = G_{MN} + B_{MN}$, we find
\begin{eqnarray}
G_{11}' &=& \frac{4}{G_{11}}
\nonumber \\
L_{1M}' &=& - 2 \, \frac{L_{1M}}{G_{11}}
\nonumber \\
L_{M1}' &=& 2 \, \frac{L_{1M}}{G_{11}}
\nonumber \\
L_{MN}' &=& L_{MN} - (-)^{MN} \frac{L_{M1} L_{1N}}{G_{11}} 
\nonumber \\
%B_{MN}' &=& 
%\nonumber \\
P^{' \alpha \hat{\beta}'} &=& \left( P^{\alpha \hat \beta} + 2 \frac{E_{1}^{~\alpha}{E_{1}^{~\hat \beta}}}{G_{11}} \right) \Gamma_{\hat \beta}^{~\hat{\beta}'}
\nonumber \\
{E'_{M}}^{\alpha} &=& Q_{M}{}^{N} {E_{N}}^{\alpha} \\
{\hat{E}'_{M}}\negmedspace {}^{~\hat{\alpha}'} &=& \hat{Q}_{M}{}^{N} {\hat{E}_{N}}^{~\hat \alpha} \Gamma_{\hat \alpha}^{~\hat{\alpha}'}
\nonumber \\
{\Omega'_{M \alpha}}^{\beta} &=& Q_{M}{}^{N} {\Omega_{N \alpha}}^{\beta}
\nonumber \\
\hat{\Omega}'_{M \hat{\alpha}'}\,^{\hat{\beta}'} &=& \hat{Q}_{M}{}^{N}  \hat \Omega_{N \hat \alpha}\,^{\hat \beta} \Gamma_{~\hat{\alpha}'}^{\hat\alpha} \Gamma_{\hat\beta}^{~\hat{\beta}'}
\nonumber \\
{C'_{\alpha}}^{\beta  \hat{\gamma}'} &=& \left( {C_{\alpha}}^{\beta \hat \gamma}  - \frac{2}{G_{11}} \Omega_{1 \alpha}{}^{\beta} \hat E_{1}^{~\hat \gamma} \right) \Gamma_{\hat\gamma}^{~\hat{\gamma}'} \nonumber \\
{{\hat{C}}_{\hat{\alpha}'}}^{' ~ \hat{\beta}'  \gamma} &=& \left(
  {\hat{C}_{\hat \alpha}}^{~\hat \beta  \gamma}  - \frac{2}{G_{11}} \hat
  \Omega_{1 \hat \alpha}{}^{\hat \beta}  E_{1}^{~\gamma} \right) \Gamma_{~\hat{\alpha}}^{\hat\alpha} \, \Gamma_{\hat\beta}^{~\hat{\beta}'} \nonumber \\
{S'}_{\alpha \hat{\gamma}'}^{\beta \hat{\delta}'} &=&
\left( S_{\alpha\hat\alpha}^{\beta\hat\beta} - {\Omega_{1\alpha}}^{\beta} \hat
\Omega_{1 \hat\alpha}{}^{\hat\beta} \right) \Gamma_{~\hat{\gamma}'}^{\hat\alpha} \,
\Gamma_{\hat \beta}^{~\hat{\delta}'} \nonumber
\end{eqnarray}
These transformations contain all fermionic corrections to the T-duality. Our
derivation of the T-duality rules is considerably more concise than
the derivations in the literature.  It can be checked that when the
background is on-shell, \emph{i.e.} it satisfies the torsion constraints, the
dual background is also on-shell. In this case the transformation rules are a
bit simpler, because the torsion constraint
$$T_{a \alpha}^{~\beta} = 0 = \hat{T}_{a \hat{\alpha}}^{~\hat{\beta}}$$
together with the condition that the fields do
not depend on $x^{1}$, implies that $\Omega_{1\alpha}^{~\beta}
=0=\hat{\Omega}_{1\hat{\alpha}}^{~\hat{\beta}}$.

The fact that an on-shell background is transformed into another on-shell background can
also be argued purely in world-sheet terms.  We need to show that the BRST
symmetry of the original action carries over to the dual action.  Since the
gauge field $A$ and the Lagrange multiplier $y$ are BRST invariant, and we
assume that the original action is invariant as well, the total gauged action
is invariant. To go to the dual theory we integrate out $x_{1}$, which is not
closed under BRST so naively we seem to break the symmetry. To show that this
is not the case, we have to perform a field redefinition before integrating
out; we shift the gauge field as follows:
\begin{eqnarray}  
A & \rightarrow & A + \frac{1}{G_{11}}  E_{1}^{\alpha} d_{\alpha} \partial x^{1} \nonumber\\ 
\bar A & \rightarrow & \bar A + \frac{1}{G_{11}} E_{1}^{\hat \alpha} \hat d_{\hat \alpha} \bar \partial x^{1} \nonumber
\end{eqnarray}
The effect of this shift in the action (\ref{gaugeact}) is to cancel the
couplings between $d$ and $\partial x^{1}$ and replace it with a coupling
between $d$ and $\partial y$. As a result, the symplectic structure on the space of fields is
modified, and it can be shown that the BRST charge (formally given by the same
expression) now leaves $x^{1}$ invariant and acts instead on $y$. It is then
safe to integrate out the isometry coordinate $x^{1}$ and the gauge potential $A$, and we arrive again at the dual action
(\ref{dualact}), having preserved the BRST invariance at each step.
 
 In order to compare our results with those of \cite{Kulik:2000nr} it is
 important to keep in mind that we are working in a different superspace. More
 precisely, Berkovits's formalism naturally gives the formulation of
 supergravity in Weyl superspace \cite{Howe:1997rf}, since the spin connection
 takes values in $spin(1,9)\times \mathbb{R}$ (corresponding to the 0 and
 2-form part of $\Omega_{M\alpha}^{~~~\beta}$ viewed as a matrix in the
 Clifford algebra; the 4-form part has to vanish for the action to be
 gauge-invariant).  While it is possible to reduce the structure group to the
 Lorentz part only, this has some consequences on the structure of the torsion
 constraints. In particular, in ordinary superspace one of the constraints
 reads
$$T_{\alpha \beta}^{~\gamma} = \left( \delta_{(\alpha}^{\gamma}
  \delta_{\beta)}^{\rho} + (\gamma_{a})_{\alpha \beta} (\gamma^{a})^{\gamma
    \rho} \right) \, \Lambda_{\rho} $$
where $\Lambda$ is the dilatino. In
Weyl superspace it is possible to set $T_{\alpha \beta}^{~\gamma} = 0$, but
the dilatino is then absorbed in the spin connection (see \cite{Tsimpis:2005vu} for a thorough 
discussion).

\section{Regularization and quantum equivalence}
We have been careful in choosing a world-sheet formalism that can be
regularized \cite{Schwarz:1992te}\cite{DeJaegher:1998pp} and quantized
\cite{Berkovits:2000fe}. In particular, the only quantum calculation we
need to perform is the Gaussian integration over the gauge field and the
coordinates. It is rigorously performed in
\cite{Schwarz:1992te}\cite{DeJaegher:1998pp}, and we summarize those
discussions.

 We introduce a generalized Hodge decomposition of the gauge field on
the world-sheet $A_a=\partial_{a} \alpha + {\epsilon_{a}}^b \partial_b
  \beta/G_{11}$.  The Gaussian integrations that one performs to obtain
either the original or the dual model are over the variables $x^1- \alpha, y$
and $\beta$.  The Gaussian integration can be regularized efficiently by
introducing either dimensional regularization as in \cite{Tseytlin:1991wr} or
a Pauli-Villars regulator field for each integration variable \cite{DeJaegher:1998pp}, with a kinetic
term determined by the quadratic terms in the fields $x^1-\alpha,y,\beta$.

The crucial observation that we make now is that in the pure spinor world-sheet
action (for a generic background), the quadratic terms are identical to the
ones in the Neveu-Schwarz Ramond formalism for a purely NS-NS background up to the fact that the coefficients are superfields
in our context, and does not affect the calculation.  We therefore conclude
that the action for the regulator fields is identical to the regulated action
in the Neveu-Schwarz Ramond formalism. Therefore, as in
\cite{DeJaegher:1998pp}, in either regularization scheme, the Gaussian
integration provides us with an equality between regularized path integrals
for the dual world-sheet quantum field theories.  Thus the equivalence of
non-linear sigma models is valid to all orders in world-sheet perturbation
theory (in the string length over the curvature radius squared), and even
non-perturbatively. To establish this it is crucial to demonstrate
that the non-local contributions to the
regulated actions on either side match, as demonstrated in 
\cite{Schwarz:1992te}.
Precisely as for purely NSNS backgrounds it can be shown that
conformal models on one side of the duality are mapped into conformal dual
models provided one shifts the dilaton, which takes into account the conformal
anomaly.  The independence of the regularization procedure on
the Ramond-Ramond backgrounds is responsible for the fact that the dilaton
shifts with an amount that depends on the Neveu-Schwarz Neveu-Schwarz
(superfield) background only.  The transformation rule is then
$$\Phi ' = \Phi - \frac12 \, \textrm{ln} G_{11} \, .$$

That provides the T-duality rule for the dilaton in the 
 final line in the action (\ref{action}). The pure spinor action is unaltered.

An important point we want to make is that we can do the full
regularization of all fields in a dimensional regularization scheme. That
scheme does not break BRST invariance of the action on both
sides of the T-duality. Thus, on-shell backgrounds are mapped onto on-shell
backgrounds, at the quantum level, in the dimensional regularization scheme.

To argue for T-duality to all order in the string coupling
$g_s$, we reason as follows. Given the
fact that on any given world-sheet (with given topology and modular parameters)
we can
demonstrate non-perturbative equivalence of the world-sheet models, it suffices
to observe that the modular integrals will be identical for type IIA and type
IIB string theories. We thus provide a proof of T-duality that is perturbative in the
string coupling, and non-perturbative in the string length over the curvature
radius, and this in any background including those with Ramond-Ramond fields.

\section{Conclusions}
The world-sheet pure spinor formalism enables us to efficiently derive the
T-duality rules for string backgrounds, as well as to generalize the proof of
equivalence of backgrounds non-perturbatively on the world-sheet and
perturbatively in target space to any non-trivial background of string theory.
Moreover, the duality holds for backgrounds that are on-shell or off-shell.
  It is interesting to
further study global aspects of the duality in backgrounds with
 Ramond-Ramond fluxes.
 T-duality on superspaces using the 
world-sheet spinor formalism also deserves investigation.
It should be possible using our techniques to analyze the important case of the $AdS_5$ background, 
that has been the subject of recent investigations (see e.g.
\cite{Alday:2007hr}\cite{Ricci:2007eq}).
\vskip .5cm

\noindent {\bf Acknowledgments} 
We thank Costas Bachas and Dimitri Tsimpis for helpful discussions. 
Our work was supported in part
by the EU under the contract MRTN-CT-2004-005104.

  \end{document}